\documentclass[11pt]{article}
\usepackage{times} 
\usepackage{amsmath} 
\usepackage{amsfonts} 
\usepackage{psfig} 
\usepackage{latexsym}
\usepackage{fullpage}
\usepackage{colordvi}

\def\MyStretch{1.1}
\def\baselinestretch{\MyStretch}

\def\eps{\varepsilon}
\def\P{{Pr}} %%%   appears in many equations  Prob

\def\Ain{A}
\def\Aout{D}
\def\Smin{S}

\def\Seff{{\cal S}^{\eps}}
\def\S{{\cal S}}
\def\T{{\cal T}}

\def\Smax{\overline{S}}
\def\deconv{\oslash}
\def\conv{*}

\newtheorem{theorem}{Theorem}
\newtheorem{lemma}{Lemma}

\begin{document}

\title{\bf A Calculus for End-to-end Statistical Service Guarantees
\thanks{This work is supported 
in part by the National Science Foundation through grants 
ANI-9730103, ECS-9875688 (CAREER), ANI-9903001, DMS-9971493, 
and ANI-0085955, and by an Alfred P. Sloan research fellowship.}
\\[10pt] 
{\small\em Technical Report: University of Virginia, CS-2001-19 (2nd revised 
version)}}

\author{ {\em Almut Burchard} $^{\dag}$ \hspace{2cm}
{\em J\"{o}rg Liebeherr} $^{\ast}$ \hspace{2cm}
{\em Stephen Patek} $^{\ast\ast}$
\\ \\ 
$^{\dag}$ Department of Mathematics\\
$^{\ast}$ Department of Computer Science\\
$^{\ast\ast}$ Department of Systems and 
Information Engineering\\
University of Virginia\\
Charlottesville, VA 22904   
}
\date{}

\maketitle

\setcounter{footnote}{0}
\thispagestyle{empty}
\def\mynormality#1{\def\baselinestretch{#1}\small\normalsize}

\mynormality{1.0}

\begin{abstract}
The deterministic network calculus offers an elegant 
framework for determining delays and backlog in a network with 
deterministic service guarantees to individual traffic flows. 
This paper addresses the problem of extending the network calculus 
to a probabilistic framework with statistical service guarantees. 
Here, the key difficulty relates to expressing, in a statistical 
setting, an  end-to-end (network) service curve as a concatenation of 
per-node service curves. 
The notion of an {\em effective service curve} is developed as a 
probabilistic bound on the service received by an individual flow. 
It is shown that per-node effective service curves can be concatenated to yield 
a network effective service curve. 
\end{abstract}


\mynormality{1.1}

\section{Introduction}

The deterministic network calculus recently evolved as a 
fundamental theory for quality of service (QoS) networks, 
and has provided powerful tools for reasoning 
about delay and backlog in a network with 
service guarantees to individual traffic flows. 
Using the notion of arrival envelopes and 
service curves \cite{Cruz95}, 
several recent works have shown that delay and backlog 
bounds can be concisely expressed in 
a min-plus algebra \cite{Cruz99,LeBoudec98,Chang98a}.

However, the deterministic  view of traffic 
generally overestimates the actual resource requirements of a flow 
and results in a low utilization of available network resources. 
This motivates the search for a statistical network calculus 
that can exploit statistical multiplexing, while preserving the 
algebraic aspects of the deterministic calculus. 
The problem of developing 
a probabilistic network calculus has  been 
the subject of several studies. Kurose \cite{Kurose92} uses the concept 
of stochastic ordering 
and obtains bounds on the distribution of delay  and  
buffer occupancy  of a flow in a network with FIFO scheduling. 
Chang \cite{chang94} presents probabilistic bounds on 
output burstiness, backlog and delays in  a network 
where the moment generating functions of arrivals are exponentially bounded.
Different bounds for stochastically bounded arrivals 
are derived by Yaron and Sidi \cite{Yaron93} 
and Starobinski and Sidi \cite{StaSi00}. 
The above results can be used to determine stochastic 
end-to-end performance bounds. 
Results on statistical end-to-end delay guarantees in a network 
have been obtained for specific scheduling algorithms, 
such as EDF \cite{Chiussi99,Chiussi2000}, and GPS \cite{ElMi99}, and a  class 
of coordinated scheduling algorithms \cite{andrews00,knightly2000b}.
Several researchers have considered probabilistic formulations of 
service curves. 
Cruz defines a probabilistic service curve which violates a given 
deterministic service curve according to a certain distribution  
\cite{Cruz96a}. 
Chang (see \cite{Book-Chang}, Chp. 7) presents exercises which 
hint at a statistical network calculus for the class of `dynamic F-servers'. 
Finally, Knightly and Qiu \cite{Knightly99b} derive  
`statistical service envelopes' as time-invariant lower bounds 
on the service received by an aggregate of flows. 

With exception of (\cite{Book-Chang}, Chp. 7), none of the cited works 
express statistical end-to-end performance bounds in a min-plus algebra, 
and it has been an open question whether a statistical network calculus 
can be developed in this setting. 
The contribution of this paper is the presentation of 
a statistical network calculus that uses the min-plus algebra 
\cite{Cruz99,LeBoudec98,Chang98a}. The advantage of using the min-plus algebra 
is that end-to-end guarantees can be expressed as a simple concatenation 
of single node guarantees, which, in turn, can be exploited to achieve 
simple probabilistic bounds. 

We define an {\em effective service curve}, which is,
with high certainty, a probabilistic bound on the service received by a single 
flow.  We will show that the main results of the deterministic network calculus 
carry over to the statistical framework we present. 
Our derivations reveal a difficulty that occurs when calculating 
probabilistic service guarantees for multiple nodes. 
We show that the problem can be overcome either by adding 
assumptions on the traffic at nodes or by modifying 
the definition of the effective service curve. 
The results in this paper are set in a continuous time model with 
fluid left-continuous 
traffic arrival functions, as is common for network delay analysis 
in the deterministic network calculus. 
A node represents a router (or switch) in a network.   
Packetization delays and other effects of discrete-sized packets, 
such as the  non-preemption of packet transmission, are ignored.  
We refer to \cite{Book-Chang} for the issues involved in  relaxing these 
assumptions for the analysis of packet networks. 
When analyzing delays in a network, all processing overhead and propagation 
delays are ignored. 
As in the deterministic network calculus, arrivals from a traffic flow 
to the network satisfy deterministic upper bounds, 
which are enforced by a deterministic regulator.

The remaining sections of this paper are structured as follows.
In Section~\ref{sec-detcalc}, we review the notation and key results 
of the deterministic network calculus. In Section~\ref{sec-statcalc} 
we introduce effective service curves and present the results for 
a statistical network calculus in terms of effective service curves. 
In Section~\ref{sec-motivate} we provide 
a discussion that motivates our revised definition of 
an effective service curve. 
In Section~\ref{sec-concl}, we present brief conclusions.

\section{Network Calculus Preliminaries}
\label{sec-detcalc}

The deterministic network calculus, which was created 
in \cite{Cruz91a,Cruz91b} and fully developed in the 
last decade, 
provides concise expressions for upper bounds on the backlog  
and delay experienced by an individual flow at one or more 
network nodes.  An attractive feature of the 
network calculus is that end-to-end bounds can often 
be easily obtained from manipulations of the 
per-node bounds. 

In this section we review some notation and 
results from the deterministic network calculus. 
This section is not a comprehensive summary 
of the network calculus and we refer to 
\cite{Cruz99,Book-LeBoudec,Book-Chang} for a complete discussion.

\subsection{Operators}

Much of the formal framework of the network calculus 
can be elegantly expressed in a min-plus algebra 
\cite{Book-Baccelli}, 
complete with convolution and deconvolution operators for functions. 
Generally, the functions in this paper are non-negative, non-decreasing, and 
left-continuous, defined over 
time intervals $[0,t]$. We assume for a given 
function $f$ that $f(t)=0$ if $t\leq 0$. 

The {\em convolution} $f \conv g$ of two functions $f$ and $g$, is defined as 
\begin{equation}
f \conv g (t) = \inf_{\tau \in [0, t]} 
\left\{ f(t-\tau) + g (\tau)\right\} \ . 
\end{equation}

The {\em deconvolution} $f \deconv g$ of two functions $f$ and $g$ is defined as 
\begin{equation}
f \deconv g (t) = \sup_{\tau \geq 0} 
\left\{ f(t+\tau) - g (\tau)\right\} \ . 
\end{equation}

For $\tau\ge 0$, the {\em impulse function} $\delta_\tau$
is defined as 
\begin{equation}
\delta_\tau (t) = \left\{ 
\begin{array}{l l}
\infty \ , & \mbox{ if } t >  \tau \ ,  \\
0  \ , &\mbox{ if } t \leq  \tau \ . 
\end{array} 
\right.
\end{equation}

If $f$ is nondecreasing, we have the formulas
\begin{eqnarray}
\label{eq-impulse}
f(t-\tau) &=&f\conv  \delta_\tau(t)\ ,\\
f(t+\tau) &=&f\deconv \delta_\tau(t)\ .
\end{eqnarray}

We refer to \cite{Book-Baccelli,Book-LeBoudec,Book-Chang} for a detailed 
discussion of the properties of the min-plus algebra and 
the properties of the convolution and deconvolution operators.

\subsection{Arrival functions and Service Curves}
 
Let us consider the traffic arrivals to a single 
network node.  The arrivals of a flow in the time 
interval  $[0,t)$ are given in terms of a function $\Ain (t)$.  
The departures of a flow from the node in the time interval 
$[0,t)$ are denoted by $\Aout (t)$, 
with $\Aout (t) \leq \Ain (t)$. The backlog 
of a flow at time $t$, denoted by $B(t)$, is given by 
\begin{equation}B(t) = \Ain(t) - \Aout(t) \ .\end{equation} 
The delay at time $t$, denoted as $W(t)$, is the 
delay experienced by an arrival which departs at time 
$t$, given by 
\begin{equation}W(t) = \inf \{ d \geq 0 \, | \; \Ain(t - d) \leq D (t) \} \ . \end{equation}
We will use $\Ain (x,y)$ and $\Aout (x,y)$ 
to denote the arrivals and departures in the time  interval   
$[x,y)$, with $\Ain (x,y)=\Ain (y)-\Ain (x)$ and 
$\Aout (x,y)=\Aout (y)-\Aout (x)$. 

We make the following assumptions on the arrival functions.
\begin{itemize}
 \item [(A1)] {\em Non-Negativity.} 
The arrivals in any interval of time are non-negative. 
That is, for any $x< y$, we have $\Ain (y) - \Ain (x)\ \geq 0$. 

\item [(A2)] {\em Upper Bound.} 
The arrivals $\Ain$ of a flow are bounded by a
subadditive\footnote{A function 
$f$  is {\em subadditive} if $f(x + y) \le f(x) + f(y)$, for all $x, y \geq 0$, 
or, equivalently, if $f(t) = f \conv f (t)$.} 
function $A^*$, 
called the {\em arrival envelope},\footnote{\label{footnote-env}
A function $E$ is called an {\em envelope} for a function $f$ if
$ f(t+\tau) - f(\tau) \leq E(t)$ for all $t,\tau \geq 0$, 
or, equivalently,  if $f(t)  \leq E  \conv f (t) \mbox{, for all }  t \geq 0$.} 
such that 
\(
\Ain (t+ \tau )  - \Ain (t) \leq A^*(\tau) 
\) 
for all $t,\tau \geq 0$. 
\end{itemize}

A {\em minimum service curve} for a flow is a 
function $\Smin$ which specifies a lower bound 
on the service given to the flow such that, for all $t \geq 0$, 
\begin{equation}
\label{eq-def-service-1}
\Aout(t) \geq \Ain * \Smin (t) \ . 
\end{equation}

A {\em maximum service curve} for a flow is a function $\Smax$ which 
specifies an upper bound on the service given to a flow such that,  
for all $t \geq 0$, 
\begin{equation} 
\label{eq-def-service-2}
\Aout (t) \leq \Ain * \Smax (t) \ . 
\end{equation}  
Minimum service curves play a larger role in the network calculus 
since they provide service guarantees. Therefore, 
we, as the related literature,  often refer to a minimum service 
curve simply as a service curve. If no maximum service curve
is explicitly given, one can use $\Smax(t)=Ct$, where $C$ is 
the link capacity.

The  following two theorems summarize some key  results of 
the deterministic network calculus.   These  results have been derived in  
\cite{Cruz99,LeBoudec98,Chang98a}. 
We follow the 
notation used in \cite{Cruz99}. 

\begin{theorem} {\bf Deterministic Calculus \cite{Cruz99,LeBoudec98,Chang98a}.}
\label{theorem-detcalc}
Given a flow with arrival envelope $A^*$ and with 
minimum service curve $\Smin$, the following hold: 

\begin{enumerate}
\item {\bf Output Envelope.} 
The function $D^* = A^* \deconv \Smin$ is an 
envelope for the departures, in the sense 
that, for all $t,\tau \geq 0$, 
\begin{equation}D^* (t) \geq D (t+\tau) - \Aout(\tau) \ . \end{equation}

\item {\bf Backlog Bound.} 
An upper bound for the backlog, denoted by $b_{max}$, 
is given by 
\begin{equation}
b_{max} = A^* \deconv \Smin (0) \ .
\end{equation} 

\item {\bf Delay Bound.} 
An upper bound for the delay, denoted by $d_{max}$, 
is given by 
\begin{equation}d_{max} = \inf \left\{ d\geq 0 \; | \;  
\forall t \ge 0 :\  A^* (t-d) \leq \Smin (t) 
\right\} \ .
\end{equation} 
\end{enumerate}
\end{theorem}

\begin{figure}[t]

\centerline{
\psfig{figure=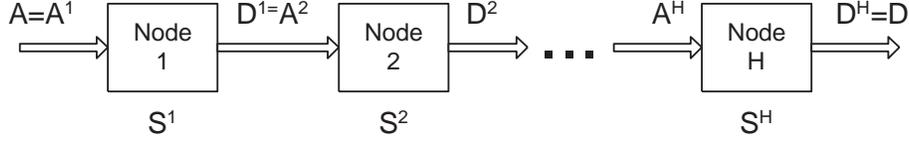,width=5in} 
}

\caption{\footnotesize Traffic of a flow through a set of $H$ nodes.
Let $\Ain^h$ and $\Aout^h$ denote the arrival and departures at the 
$h$-th node, with $\Ain^1 = A$, $A^h = D^{h-1}$ 
for $h = 2, \ldots, H$ and $D^H=D$.} 
\label{fig-net}

\end{figure}

The next theorem states that the service curves of a flow at the nodes on its route 
can be concatenated to define a network service curve, which expresses 
service guarantees offered to the flow by the network as a whole. 

\begin{theorem} {\bf Concatenation of Deterministic Network Service Curves 
\cite{Cruz99,LeBoudec98,Chang98a}.}
\label{theorem-detnet}
Suppose a flow passes  through $H$ nodes in series, 
as shown in Figure~\ref{fig-net},
and suppose the flow is offered minimum and maximum service curves
$\Smin^h$  and $\Smax^h$, respectively, at 
each node $h=1,\ldots,H$.
Then, the sequence of nodes provides minimum and  maximum 
service curves  $\Smin^{net}$ and $\Smax^{net}$, 
which are given by 
\begin{eqnarray}
\label{eq-def-Snetmin}
\Smin^{net} & = & \Smin^1 \conv 
\Smin^2 \conv \ldots \conv \Smin^H \ , \\ 
\label{eq-def-Snetmax}
\Smax^{net} & = & \Smax^1 \conv \Smax^2 \conv \ldots \conv \Smax^H   \ . 
\end{eqnarray}
\end{theorem}
$\Smin^{net}$ and $\Smax^{net}$ will be referred to as 
{\em network service curves}, and 
Eqs.~(\ref{eq-def-Snetmin})--(\ref{eq-def-Snetmax}) will be called the 
{\em  concatenation formulas}.

With Theorems~\ref{theorem-detcalc} and ~\ref{theorem-detnet} network service curves can 
be used to determine bounds on delay and backlog for individual
flows in a network.  
There are many additional properties and refinements that have 
been derived for the deterministic calculus. However, 
in this paper we will concern ourselves 
only with the results above.

\section{Statistical Network Calculus}
\label{sec-statcalc}

We now approach 
the network calculus in a probabilistic framework.
Arrivals and departures from a flow to the network
in the time interval  $[0,t)$ are described by
random processes $\Ain(t)$ and $\Aout(t)$
satisfying assumptions (A1) and (A2).
The random processes are defined over  
an underlying joint probability space that we suppress 
in our notation. 
The statistical network calculus makes service 
guarantees for individual flows, where each 
flow is allocated a probabilistic service in the 
form of an `effective service curve'.

Given a flow with arrival process $A$, 
a {\em (minimum) effective service curve} 
is a nonnegative function $\Seff$ that satisfies for all $t>0$,
\begin{equation}
\label{eq-eff-servicecurve}
Pr \bigl\{
\Aout (t) \geq \Ain*\Seff(t) \bigr\} \geq 1 - \eps  \  .
\end{equation}
 
Note that the effective service curve is a 
non-random function.  We omit the corresponding 
definition of a {\em maximum effective service curve}. 

The following theorem is a probabilistic counterpart 
to Theorem~\ref{theorem-detcalc}.

\begin{theorem} {\bf Statistical Calculus.}
\label{theorem-statcalc}
Given a flow with arrival process $A$ satisfying assumptions (A1)--(A2),
and given an effective service curve $\Seff$,
the following hold: 
\begin{enumerate}
\item {\bf Output Envelope.} 
The function $A^\ast \deconv \Seff$ is a
probabilistic bound for the departures on $[0,t]$, in the sense that, 
for all $t,\tau >0$, 
\begin{equation}
Pr \left\{ \Aout (t, t+\tau) \le A^\ast \deconv \Seff (\tau) \right\} 
\ge 1-\eps  \ . 
\end{equation} 
\item {\bf Backlog Bound.} 
A probabilistic bound for  the backlog is given by 
$b_{max} = A^* \deconv \Seff (0)$, 
in the sense that, for all $t>0$,
\begin{equation}
Pr \left\{ B (t) \leq  b_{max} 
\right\} \ge 1-\eps   \  . 
\end{equation} 

\item {\bf Delay Bound.} 
A probabilistic bound for  the delay  is given by, 
\begin{equation}d_{max} = \inf \left\{ d\geq 0 \; | \;  
\forall t \ge 0 :\  A^* (t - d) \leq \Seff (t) 
\right\} \ , 
\label{eq-def-dmax}
\end{equation} 
in the sense that, for all $t>0$,
\begin{equation}
Pr \left\{ W(t) \leq d_{max} \right\} \ge 1-\eps \ . 
\end{equation} 
\end{enumerate}
\end{theorem}

By setting $\eps = 0$ in Theorem~\ref{theorem-statcalc}, we 
can recover the bounds of Theorem~\ref{theorem-detcalc} with probability one. 

\medskip
\noindent{\bf Proof.} \ 
The proof uses on several occasions  that the inequality
\begin{equation}
\label{eq-detcalc-proof1-0}
\Ain (t+\tau) - A*g(t) \le A^*\deconv g(\tau)\ \
\end{equation}
holds for any nonnegative function $g$ and
for all $t,\tau\ge 0$. To see this inequality, we compute
\begin{eqnarray}
\Ain (t+\tau) - A*g(t) 
&=& \Ain (t+\tau) - \inf_{x \in [0, t]} \bigl\{ \Ain (t - x) + g (x ) \bigr\}
\label{eq-detcalc-proof1-2}\\
& =  & \sup_{x \in [0, t]} \bigl\{ \Ain (t - x, t +\tau) -  g (x)\bigr\}
\label{eq-detcalc-proof1-3}\\
& \leq & \sup_{x \geq 0} \bigl\{ A^* (\tau + x) -  g (x)\bigr\}
\label{eq-detcalc-proof1-4}\\
& = & A^* \deconv g (\tau)   \ . 
\label{eq-detcalc-proof1-5}
\end{eqnarray}
Eqn.~(\ref{eq-detcalc-proof1-2}) expands the convolution operator. 
Eqn.~(\ref{eq-detcalc-proof1-3})  takes $\Ain (t+\tau)$ 
inside the infimum and uses that 
$\Ain (t +\tau) -  \Ain (t - x)= \Ain (t - x, t +\tau)$. 
Eqn.~(\ref{eq-detcalc-proof1-4}) uses that $A(t-x,t+\tau)\le A^*(x+\tau)$
for all $x\le t$ by definition of an arrival envelope, and extends
the range of the supremum.   Finally, Eqn.~(\ref{eq-detcalc-proof1-5}) 
uses the definition of the deconvolution operator. 

\medskip 
\noindent 
{\em 1. Proof of the Output Bound. }
For any fixed $t,\tau > 0$,  we have 
\begin{eqnarray}
1 - \eps 
& \leq &  \P  \left\{  \Aout (t) \ge \Ain  \conv  \Seff (t) \right\} 
\label{eq-statcalc-proofI-0}\\
& = &  \P  \left\{  \Aout (t,t+\tau) \le \Aout (t+\tau) - 
                 \Ain  \conv  \Seff (t) \right\} 
\label{eq-statcalc-proofI-1}\\
& \leq &  
\P  \left\{  
\Aout (t,t+\tau) \le \Ain (t+\tau) - \Ain  \conv  \Seff (t) 
\right\} 
\label{eq-statcalc-proofI-2} \\
& \le  & \P  \left\{  \Aout (t,t+\tau) \le 
A^* \deconv \Seff (\tau)  
\right\}  \ . 
\label{eq-statcalc-proofI-3}
\end{eqnarray}
Eqn.~(\ref{eq-statcalc-proofI-0}) holds by the definition
of the effective service curve $\Seff$.
Eqn.~(\ref{eq-statcalc-proofI-1}) uses 
that $\Aout(t,t+\tau)=\Aout(t+\tau)-\Aout(t)$.
Eqn.~(\ref{eq-statcalc-proofI-2}) uses that 
departures in $[0,t)$ cannot exceed arrivals, 
that is, $\Aout (t) \leq \Ain (t)$ for all $t \geq 0$. 
Finally, Eqn.~(\ref{eq-statcalc-proofI-3})
uses that $ \Ain (t+\tau) - \Ain  \conv  \Seff (t) \le
A^*\deconv \Seff(\tau)$ by Eqn.~(\ref{eq-detcalc-proof1-0}).

\medskip 
\noindent 
{\em 2. Proof of the Backlog Bound.} 
Since $B(t)=\Ain(t)-\Aout(t)$ and with the definition
of the effective service curve, we can write
\begin{eqnarray}
1 - \eps 
& \leq &  \P  \left\{  D(t) \ge \Ain\conv\Seff(t)\right\} 
\label{eq-statcalc-proofII-1}\\
& = &  \P  \left\{  B(t) \le A(t)- \Ain\conv\Seff(t)\right\}
\label{eq-statcalc-proofII-2}\\
& \le  & \P  \left\{  B(t) \le A^* \deconv \Seff (0)  \right\}  \ . 
\label{eq-statcalc-proofII-8}
\end{eqnarray}
Eqn.~(\ref{eq-statcalc-proofII-1}) holds
by definition of the effective service curve.
Eqn.~(\ref{eq-statcalc-proofII-2}) uses that 
$B (t)  =  \Ain (t) - \Aout (t)$,
and Eqn.~(\ref{eq-statcalc-proofII-8})
uses that $A(t)- \Ain\conv\Seff(t)\le A^*\deconv \Seff(0)$
by Eqn.~(\ref{eq-detcalc-proof1-0}).

\medskip 
\noindent 
{\em 3. Proof of the Delay Bound.}  
The delay bound is proven by 
estimating the probability that the output $\Aout(t)$ exceeds 
the arrivals $A(t - d_{max})$. 
\begin{eqnarray}
1 - \eps & \leq &  \P  \left\{  \Aout(t) \geq \Ain \conv  \Seff (t) \right\} 
\label{eq-statcalc-proofIII-1}\\
& \leq &  \P  \left\{  \Aout(t) \geq \Ain \conv  
          ( A^* \conv \delta_{d_{max}} ) (t)  \right\} 
\label{eq-statcalc-proofIII-2}\\
& \leq &  \P  \left\{  \Aout(t) \geq (\Ain \conv  
           A^* )\conv \delta_{d_{max}} ) (t)  \right\} 
\label{eq-statcalc-proofIII-2a}\\
& \le &  \P  \left\{  \Aout(t) \geq \Ain (t-  d_{max})   \right\}  \ . 
\label{eq-statcalc-proofIII-3}
\end{eqnarray}
Eqn.~(\ref{eq-statcalc-proofIII-1}) uses the definition of the 
effective service curve $\Seff$.  
Eqn.~(\ref{eq-statcalc-proofIII-2}) uses the definition
of the impulse function in Eqn.~(\ref{eq-impulse})
and the definition of $d_{max}$ in Eqn.~(\ref{eq-def-dmax}).
Eqn.~(\ref{eq-statcalc-proofIII-2a}) follows from
the associativity of the convolution, and
Eqn.~(\ref{eq-statcalc-proofIII-2a}) uses the definition
of an arrival envelope.

\hfill$\Box$

\bigskip
A probabilistic counterpart to Theorem~\ref{theorem-detnet} 
can be formulated as follows.
\begin{theorem} {\bf Concatenation of 
Effective Service Curves.} 
\label{theorem-statnet-intervals-x}
Consider a flow that passes through $H$ network nodes in series,
as shown in Figure~\ref{fig-net}. Assume that effective service curves 
are given by nondecreasing functions
$\S^{h,\eps}$ at each node  ($h=1,\dots , H$).
Then, for any $t \geq 0$, 
\begin{equation}
\label{eq-statnet-intervals-x}
Pr\Bigl\{ \Aout(t)\ge 
\Ain\conv \bigl( \S^{1,\eps} \conv \dots \conv \S^{H,\eps}
    \conv \delta_{(H-1)a}\bigr) (t) \Bigr\} \ge 1-\eps\left(1+ 
                                              (H\!-\!1) \frac{t}{a}\right)\ ,
\end{equation}
where  $a>0$ is an arbitrary parameter.
\end{theorem}

Again, we can can recover the deterministic result from Theorem~\ref{theorem-detnet}. 
By setting $\eps = 0$, the results in 
Eqn.~(\ref{eq-statnet-intervals-x}) hold 
with probability one. Then by letting $a \longrightarrow 0$, we obtain 
Theorem~\ref{theorem-detnet} almost surely. 

\medskip
\noindent{\bf Proof.} \ 
We proceed in three steps. In the first step, we modify the effective 
service curve to give lower bounds on the departures simultaneously
for all times in the entire interval $[0,t]$.
In the second step, we perform a deterministic calculation.
The proof concludes with a simple probabilistic estimate.

\medskip\noindent {\em Step 1: Uniform probabilistic bound
on $[0,t]$.\ } 
Suppose that $S^\eps$ is a nondecreasing effective service  curve, that is
\begin{equation}
\label{eq-stronger-1}
\forall x \in [0,t]:\quad
Pr \bigl\{ \Aout(x)\ge \Ain\conv\S^\eps(x) \bigr\} \ge 1-\eps\  .
\end{equation}
We will show that then, for any choice of $a>0$,
\begin{equation}
\label{eq-stronger-2}
Pr \bigl\{ \forall x \in [0,t]: 
\Aout(x)\ge \Ain\conv\S^\eps(x-a) \bigr\} \ge 1-\eps \frac{t}{a}\ .
\end{equation}
To see this, fix $a>0$, set $ x_j=ja$, and consider the events
\begin{equation}
E_j=  \bigl\{ \Aout(x_j)\ge \Ain\conv\S^\eps(x_j)\bigr\}\ ,
\quad j = 1, \ldots , n-1 , 
\end{equation}
where $n=\lceil \ell/a\rceil$ is the smallest
integer no larger than $t/a$. 
Let $x\in [0,t]$ be arbitrary, and let $j$ the largest integer
with $x_j\le x$, so that
$x-x_j\le a$. If $E_j$ occurs, then
\begin{eqnarray}
\label{eq-stronger-2a}
\Aout(x) \ge \Aout(x_j)  \ge \Ain \conv\S^{\eps}(x_j) \  
\ge \Ain \conv\S^{\eps}(x-a)\ ,
\end{eqnarray}
where we have used the fact  that $\S^{\eps}$ is nondecreasing in 
the last step. It follows that
\begin{eqnarray}
\label{eq-stronger-3}
Pr \bigl\{ \forall x \in [0,t]: 
\Aout(x)\ge \Ain\conv\S^\eps(x-a) \bigr\} 
&\ge& Pr \bigl\{ \forall j = 1, \ldots,  n:\ 
\Aout(x_j)\ge \Ain\conv\S^\eps(x_j) \bigr\} \\
&=& Pr\Bigl\{ \bigcap_{0<j\le n} E_j\Bigr\}\\
&\ge& 1-n\eps\ ,
\label{eq-stronger-4}
\end{eqnarray}
which proves Eqn.~(\ref{eq-stronger-2}).
Thus, the assumptions of the theorem imply that
\begin{equation}
Pr \left\{ \begin{array}{lll}
\forall x \in [0,t]:
& \Aout^{h}(x) \ge \Ain^h \conv \S^{h,\eps}\conv \delta_(x) \ , \quad& h<H \\
& \Aout^{H}(t) \ge \Ain^h \conv\S^{H,\eps}(t)\ , \quad& h=H
\end{array}\right\}  \ge 1- \eps\left(1+ (H-1)\frac{t}{a}\right)\ .
\label{eq-proof-statnet-x-0}
\end{equation}

\medskip\noindent{\em Step 2: A deterministic argument.\ } 
Suppose that, for a particular sample path, 
\begin{equation}
\label{eq-proof-statnet-x-1} 
\left\{
\begin{array}{lll}
\forall x \in [0,t]:
& \Aout^{h}(x) \ge \Ain^h \conv \S^{h,\eps}\conv \delta_(x) \ , \quad& h<H\ ,\\
& \Aout^{H}(t) \ge \Ain^h \conv\S^{H,\eps}(t)\ , \quad& h=H\ .
\end{array}\right.
\end{equation}
Inserting the first line of Eqn.~(\ref{eq-proof-statnet-x-1})
with $h=H-1$ into the second line yields 
\begin{eqnarray}
\label{eq-proof-statnet-x-2}
\Aout^H(t) 
&\ge& \inf_{x \in [0, t]} \Bigl\{ \inf_{y \in [0, x]} 
\bigl\{ \Ain^{H-1}(t-x-y) + 
        \bigl(\S^{H-1,\eps}\conv\delta_a\bigr)(y) \bigr\} + 
       \S^{H,\eps}(x) \Bigr\}\\
&=& \Ain \conv \bigl(\S^{H-1,\eps} \conv \S^{H,\eps}\conv 
        \delta_{a}\bigr) (t)\ .
\end{eqnarray}
An induction over the number of nodes shows that
Eqn.~(\ref{eq-proof-statnet-x-1}) implies
that $\Ain=\Ain^1$ and $\Aout=\Aout^h$ satisfy
\begin{equation}
\label{eq-proof-statnet-x-5}
\Aout(t) \ge \Ain \conv \bigl(\S^{1,\eps} \conv 
\dots \conv \S^{H,\eps}\conv \delta_{(H\!-\!1)a}\bigr) (t)  \ .
\end{equation}

\medskip\noindent {\em Step 3: Conclusion. \ } We estimate 
\begin{eqnarray}
&& Pr \bigl\{\Aout(t) \ge \Ain \conv \bigl(\S^{1,\eps} \conv 
     \dots \conv \S^{H,\eps}\conv \delta_{(H\!-\!1)a}\bigr) (t)  \  
                              \bigr\} \\
&&\qquad \ge 
Pr \{ \mbox{\ Eqn.~(\ref{eq-proof-statnet-x-1}) is satisfied\ }\}\\ 
&&\qquad \ge 1-\eps \left( 1+ (H\!-\!1)\eps \frac{t}{a}\right)\ .
\end{eqnarray}
The first inequality follows from the fact that 
Eqn.~(\ref{eq-proof-statnet-x-1}) implies
Eqn.~(\ref{eq-proof-statnet-x-5}). The 
second inequality merely uses Eqn.~(\ref{eq-proof-statnet-x-0}).
\hfill$\Box$

\bigskip 
Since the bound in Eqn.~(\ref{eq-statnet-intervals-x})
deteriorates as $t$ becomes large, Theorem~\ref{theorem-statnet-intervals-x} 
is of limited practical value. 
To explain why Eqn.~(\ref{eq-statnet-intervals-x}) 
deteriorates, consider a 
network as shown in Figure~\ref{fig-net}, with $H=2$ nodes. An effective
service curve $\S^{2,\eps}$  in the sense of 
Eqn.~(\ref{eq-eff-servicecurve}) at the second 
node guarantees that, for any given time $t$,
the departures from this node
are with high probability bounded below by
\begin{equation}
\label{eq-tech-issue-1}
\Aout^2(t)\ge \Ain^2\conv \S^{2,\eps}(t) =
\inf_{\tau\in [0,t]} \bigl\{ \Ain^2(t-\tau)+\S^{2,\eps}(\tau) \bigr\} \ .
\end{equation}
Suppose that the infimum in Eqn.~(\ref{eq-tech-issue-1})
is assumed at some value $\hat\tau \le t$. Since
the departures from the first node are random, even if the 
arrivals to the first node satisfy the deterministic  bound $A^*$,
$\hat \tau$ is a random variable. 
An effective service curve $\S^{1,\eps}$
at the first node guarantees that for any arbitrary but fixed 
time $x$, the arrivals $\Ain^2(x)=\Aout^1(x)$
to the second node are with
high probability bounded below by
\begin{equation}
\label{eq-tech-issue-2}
\Aout^1(x)\ge \Ain^1\conv \S^{1,\eps}(x)\ . 
\end{equation}
Since $\hat \tau$ is a random variable, 
we cannot simply evaluate 
Eqn.~(\ref{eq-tech-issue-2}) for $x = t - \hat \tau$ and use the resulting 
bound in Eqn.~(\ref{eq-tech-issue-1}). 
Furthermore, there is, a priori, no 
time-independent bound on the distribution of $\hat \tau$.
Note that the  above issue does not arise in the deterministic calculus, 
since deterministic service curves make service guarantees
that hold for all values of $x$. 

We conclude that, in a probabilistic setting, 
additional assumptions 
are required to establish time-independent bounds 
on the range of the infimum, and, in that way, 
obtain probabilistic network service curves that do 
not deteriorate with time.
One example of such an assumption is to add the
condition that
\begin{equation}
\label{eq-eff-servicecurve-T}
Pr \left\{ \Aout (t) \geq 
     \inf_{x\in [0,T]} \bigl\{\Ain(t-x) +\Seff(x) \bigr\} 
\right\} 
\geq 1 - \eps  \  .
\end{equation}
This condition imposes a limit on the range of the convolution.
The condition can be satisfied 
for a given effective service curve $\Seff$ and arrival 
envelope $A^*$ by choosing $T$ such that $A^*(T)\le\Seff(T)$,
which guarantees that
\begin{equation}
\Ain\conv \Seff(t) = 
\inf_{x \in [0,T]} \bigl\{ \Ain(t-x)+\Seff(x)\bigr\}\ .
\end{equation}

\begin{theorem} 
\label{theorem-statnet-intervals-w}
Assume that all hypotheses of
Theorem~\ref{theorem-statnet-intervals-x} are 
satisfied, and additionally,
that there exists a number $T\ge 0$ such
that $\Ain^h$ and $\S^{h,\eps}$ satisfy
Eqn.~(\ref{eq-eff-servicecurve-T}) 
for $h=1,\dots , H$.  Then, for any choice of $a>0$,
\begin{equation}
\label{eq-eff-concat-1}
\S^{net,{\eps'}} = \S^{1,\eps} \conv \dots \conv \S^{H,\eps}
    \conv \delta_{(H\!-\!1)a}
\end{equation}
is an effective network service curve, with violation 
probability bounded by
\begin{equation}
\label{eq-eff-concat-1a}
\eps' \le\  H \eps \left(1 + (H\!-\!1) \frac{T+a}{2a}\right)\ .
\end{equation}
More precisely, $\S^{net,\eps'}$  satisfies
\begin{equation}
\label{eq-eff-concat-2}
Pr \left\{ \Aout (t) \geq \inf_{x\in [0,H(T+a)]} 
         \bigl\{\Ain(t-x)+ \S^{net,{\eps'}}(x) \bigr\} 
     \right\}
\geq 1 - \eps'\ .
\end{equation}
\end{theorem}
The bounds of this network service curve deteriorate with the 
number of nodes $H$, but, different from 
Theorem~\ref{theorem-statnet-intervals-x},  
the bounds are not dependent on $t$. Rather 
the bounds depend on a time scale $T$ as used in 
Eqn.~(\ref{eq-eff-servicecurve-T}).
A key issue, which is not addressed in this paper, relates 
to establishing $T$ for an arbitrary node in the network. 

\noindent{\bf Proof.} \  The proof is analogous
to the proof of Theorem~\ref{theorem-statnet-intervals-x},
and proceeds in the same three steps.

\medskip\noindent {\em Step 1: Uniform probabilistic bounds
on intervals of length $\ell$.\ \ } 
Suppose that $\Seff$ is a nondecreasing effective
service  curve satisfying Eqn.~(\ref{eq-eff-servicecurve-T}),
and let $\ell>0$. 
Fix $a>0$, set $ x_j=t-\ell+ja$, and consider the events
\begin{equation}
E_j=  \left\{ \Aout(x_j)\ge \inf_{y\in [0,T]} 
       \bigl\{ \Ain(x_j-y)+\S^\eps(y)\bigr\}\right\}\ ,
\quad j = 0, \ldots , n-1 , 
\end{equation}
where $n=\lceil \ell/a\rceil$. If $x\in [x_j,x_{j+1})$
and the event $E_j$ occurs, then a computation analogous
to Eqn.~(\ref{eq-stronger-2a}) shows that
\begin{equation}
\Aout(x) \ge  \inf_{y\in [0,T+a]} \bigl\{ \Ain(x-y)+ 
 \bigl(\S^\eps\conv\delta_a(y)\bigr)\bigr\}\ .
\end{equation}
We conclude as in Eqs.~(\ref{eq-stronger-3})-(\ref{eq-stronger-4}) that
\begin{eqnarray}
Pr \left\{ \forall x \in [t-\ell,t]: 
\Aout(x)\ge \ \inf_{y\in [0,T+a]} \bigl\{\Ain(x-y) + 
         \bigl(\Seff\conv\delta_a\bigr)(y) \bigr\}\right\} 
&\ge & Pr\Bigl\{ \bigcap_{0\le j< n} E_j\Bigr\}\\
&\ge &  1-\eps\left\lceil\frac{\ell}{a}\right\rceil\ .
\end{eqnarray}

\medskip\noindent{\em Step 2: A deterministic argument.\ } 
A computation analogous to 
Eqs.~(\ref{eq-proof-statnet-x-2})-(\ref{eq-proof-statnet-x-5}) shows that
\begin{equation}
\label{eq-proof-concat-3} 
\left\{ \begin{array}{lll}
\forall x\in [t-(H\!-\!h),t]: 
&\Aout^{h}(x) \ge \displaystyle{\inf_{y\in [0,T+a]}}
\bigl\{\Ain^h (x-y) + \bigl(\S^{h,\eps}\conv\delta_a\bigr)(y)
\bigr\}
\quad & h<H,\\
&\Aout^{H}(t) \ge \displaystyle{\inf_{y\in [0, T+a]}}
\bigl\{\Ain^H (t-y) + \S^{H,\eps}(y)
    \bigr\} 
    & h=H\ ,
\end{array}\right.
\end{equation}
implies
\begin{equation}
\label{eq-proof-concat-7}
\Aout(t) \ge \inf_{x \in [0,H(T+a)]} \{\Ain (t-x) + 
     \S^{net,{\eps'}} (x) \}  \ .
\end{equation}

\medskip\noindent {\em Step 3: Conclusion. \ }
Combining Steps 1 and 2, we obtain
\begin{eqnarray}
Pr \bigl\{\Aout(t) 
\ge  \inf_{x\in [0,H(T+a)]} \{\Ain (t-x) + 
            \S^{net,{\eps'}}(x) \} \bigr\} 
&\ge &
Pr \bigl\{ \mbox{Eqn.~(\ref{eq-proof-concat-3}) is satisfied} \bigr\}
\label{eq-proof-concat-8}\\
&\ge& 
1- \eps\left(1+ \sum_{h=1}^{H-1} 
       \left\lceil\frac{(H\!-\!h)T}{a}\right\rceil\right) 
\label{eq-proof-concat-9}\\
&\ge& 1- \eps'\ .
 \end{eqnarray}
Here, Eqn.~(\ref{eq-proof-concat-8}) follows from Step 2, and
Eqn.~(\ref{eq-proof-concat-9}) follows from  the assumptions
by choosing $\ell_h=(H\!-\!h \!+\!1) $ for $h=1,\dots, H-1$
in Step 1.
\hfill$\Box$

\section{Statistical Calculus with Adaptive 
Service Guarantees} 
\label{sec-motivate} 

We next define a class of effective service curves where 
the range of the infimum is 
bounded independently of time, and then give conditions 
under which these service curves are also effective
service curves in the sense of Eqn.~(\ref{eq-eff-servicecurve}).
The resulting effective service curves are valid without 
adding assumptions on a specific 
arrival distribution or service discipline. 
Within this context, we obtain an 
effective network service curve, where the convolution 
formula has a similarly simple form as in the deterministic network 
calculus.

\subsection{(Deterministic) Adaptive Service Curves}

\begin{figure}[t]

\centerline{
\psfig{figure=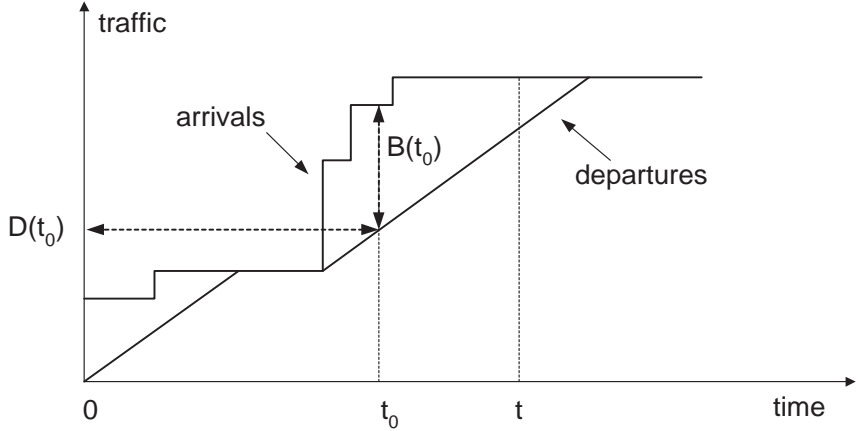,width=5in} 
}

\caption{\footnotesize Illustration for the modified convolution operator.
The operator $\conv_{t_0}$ uses the backlog at time $t_0$ and 
the arrivals in the interval $[t_0, t]$. }
\label{fig-delay_backlog}

\end{figure}

We define a modified convolution operator by setting, for any $t_0\le t$,
\begin{equation}
\label{eq-def-conv-t0}
\Ain \conv_{t_0}g(t) 
= \min \Bigl\{ g(t-t_0), B(t_0)+ 
        \inf_{\tau\le t-t_0} \bigl\{ \Ain(t_0,t-\tau)+g(\tau)\bigr\}\Bigr\}\ .
\end{equation}
The essential property of this modified operator is that 
the range over which the infimum is taken is limited to the 
interval $[t_0, t]$.   Note that the function 
$\Ain \conv_{t_0}g(t) $ depends on the backlog at time $t_0$ as 
well as on the arrivals in the interval $[t_0,t]$. 
It can be written equivalently as
\begin{equation}
\label{eq-def-conv-t0-alt}
\Ain \conv_{t_0}g(t) 
= \min \Bigl\{ g(t-t_0), 
   \inf_{\tau\le t-t_0} \bigl\{ \Ain(t-\tau)+g(\tau)\bigr\}-\Aout(t_0)\Bigr\}\ .
\end{equation}
%The operator $\conv_{t_0}g$ is causal,
%since it is determined by the arrivals and departures
%up to time $t$.  
The usual  convolution operator is recovered 
by setting $t_0=0$. 

We now reconsider the definition of a service curve in a deterministic regime. 
We introduce a revised definition of a (deterministic) service curve,
which is presented in \cite{Cruz99,Cruz96b}, and is 
referred to as {\em adaptive service curve} in \cite{Book-LeBoudec}.
A {\em (minimum) adaptive service curve} 
is defined as a function $\Smin$ which specifies a lower bound 
on the service given to a flow such that, for all $t,t_0 \geq 0$, 
with $t_0 \leq t$, 
\begin{equation}
\Aout(t_0,t) \geq \Ain\conv_{t_0}\Smin(t)\ .
\label{eq-def-intervalservice}
\end{equation}
A maximum adaptive service curve can be defined accordingly.\footnote{We note 
that the adaptive service curve in \cite{Book-LeBoudec} is more general and is defined 
using 
$\Aout(t_0,t) \geq  \min \Bigl\{ f(t-t_0), B(t_0)+ 
        \inf_{\tau\le t-t_0} \bigl\{ \Ain(t_0,t-\tau)+g(\tau)\bigr\}\Bigr\}$. 
In our context we set $f = g$.}
Eqn.~(\ref{eq-def-intervalservice}) is equivalent to requiring
that $\Smin$ satisfies Eqn.~(\ref{eq-def-service-1})
for the time-shifted arrivals and departures
\begin{equation}
\label{eq-timeshift}
\tilde \Ain(x)=B(t_0)+\Ain(t_0,t_0+x)\ ,\quad
\tilde\Aout(x)=\Aout(t_0,t_0+x)\ .
\end{equation}
Figure~\ref{fig-delay_backlog}  illustrates the time-shifted arrivals.
Many service curves with applications in 
packet networks, such as shapers, schedulers with delay guarantees, 
and rate-controlled schedulers such as GPS, can be expressed in 
terms of adaptive service curves. 
By setting $t_0 = 0$, one can see that each adaptive service curve 
is a service curve. However, the converse does not 
hold~\cite{Book-LeBoudec}. 

We next define a {\em (minimum) 
$\ell$-adaptive service curve}, denoted by $\Smin^\ell$, 
as a function for which Eqn.~(\ref{eq-def-intervalservice})
is satisfied whenever $t - t_0 \leq \ell$.
If $\ell = \infty$, we obtain an {\em adaptive service curve},
and drop the superscript in the notation.  
The difference between a service curve according to 
Eqn.~(\ref{eq-def-service-1})  and 
an $\ell$-adaptive service curve is that the former involves 
arrivals over the entire interval $[0,t]$, while the latter 
uses information about arrivals and departures in
intervals $[t_0,t]$ whose length does not depend on $t$.
Performing a time shift as in Eqn.~(\ref{eq-timeshift}) 
and applying Theorem~\ref{theorem-detnet} shows 
that the convolution of $\ell$-adaptive service curves 
yields an $\ell$-adaptive network service curve.

The following lemma shows that for $\ell$ sufficiently large,
but finite, an $\ell$-adaptive service curve is 
a service curve in the sense  of Eqn.~(\ref{eq-def-service-1}). 
In particular,  the conclusions of 
Theorems~\ref{theorem-detcalc} and \ref{theorem-detnet}
hold for such service curves.

\begin{lemma}  
Suppose that the arrival function $A$ of a flow
has arrival envelope $A^*$. Let $\Smin^\ell$ be an 
$\ell$-adaptive service curve.  If 
\begin{equation}
\label{eq-def-T0}
\exists t \in [0,\ell]:  \ \ A^*(t)\le \Smin^\ell(t) \ , 
\end{equation} 
then $\Smin = \Smin^\ell$ is an adaptive   
service curve for intervals of arbitrary length.
In particular, $\Smin$ satisfies Eqn.~(\ref{eq-def-service-1})
for all $t\ge 0$.
\label{lemma-sufficient}
\end{lemma}
The proof of the lemma is given in the appendix.

\subsection{Effective Adaptive Service Curves}

Next we introduce a probabilistic version of the 
$\ell$-adaptive service curve. We define an 
{\em effective $\ell$-adaptive service curve}
to be a nonnegative function  $\S^{\ell,\eps}$ such that 
\begin{equation}
Pr\bigl\{ \Aout(t_0,t) \geq \Ain\conv_{t_0}\S^{\ell,\eps} (t)
          \bigr\}\ \ge 1-\eps\  \ 
\label{eq-eff-intervalservice}
\end{equation}
for all $t_0,t\ge 0$ with $t-t_0\le \ell$. 
If $\ell = \infty$, we call the resulting function an 
{\em effective adaptive service curve}, and drop the superscript. 
Note that the infimum in the convolution 
on the right hand side of Eqn.~(\ref{eq-eff-intervalservice})
ranges over an interval of length at most $\ell$. 
With this bound on the range of the infimum, 
we derive the following effective network service curve. 
Technically, $\ell$ plays a similar role as the bound
$T$ on the range of the convolution in Eqn.~(\ref{eq-eff-servicecurve-T}).

\begin{theorem}\ {\bf Concatenation of Effective $\ell$-Adaptive 
Service Curves.} 
\label{theorem-statnet-intervals-y}
Consider a flow passing
through nodes numbered $h=1,\dots H$, and assume
that, at each node, an  effective $\ell$-adaptive service curve
is given by a nondecreasing function
$\S^{h,\ell,\eps_h}$. Then the function
\begin{equation}
\S^{net, \ell, \eps'} = \S^{1,\ell, \eps} \conv \ldots
                \conv \S^{H,\ell, \eps} * \delta_{(H\!-\!1)a}(t)\
\end{equation}
is an effective $\ell$-adaptive network service curve for 
any choice of $a>0$, with violation probability bounded by 
\begin{equation}
\eps' \le \eps\left( 1+ (H\!-\!1)\left\lceil \frac{\ell}{a}
\right\rceil\right)\ .
\end{equation}
\end{theorem}

\medskip
\noindent{\bf Proof.} \ 
We need to show that, for  any $t_0, t$ with $t-t_0\le \ell$
and any choice of the parameter $a$, we have
\begin{equation}
Pr \bigl\{
\Aout(t_0,t)\ge \Ain\conv_{t_0}\S^{net,\ell, H\eps\ell/a}(t) \bigr\}\ge 
1-\eps' \ .
\end{equation}
Performing a time shift as in Eqn.~(\ref{eq-timeshift}),
we may assume without loss of generality that $t_0=0$
and $t\in [0,\ell]$. The claim now follows immediately from
Theorem~\ref{theorem-statnet-intervals-x}.

\hfill$\Box$

\bigskip 
Even though the concatenation formula in 
Theorem~\ref{theorem-statnet-intervals-y} 
results in a significant improvement over 
Theorem~\ref{theorem-statnet-intervals-x}, 
a drawback of Theorem~\ref{theorem-statnet-intervals-y} 
is that the construction of the network service curve 
results in a degradation of the violation 
probability $\eps'$ and introduces a 
time shift $\delta_{(H\!-\!1)a}$, which 
grow significantly when $\ell$ and $H$ become large.
%The conversion of an effective $\ell$-adaptive to an 
%effective adaptive service curve in Lemma~\ref{lemma-eff-sufficient-y}
%introduces additional losses.
To avoid this successive degradation of the service guarantees,
we further strengthen the effective service curve. 
We define a {\em strong effective adaptive service
curve} for intervals of length $\ell$
to be a function $\T^{\ell,\eps}$
which satisfies for any interval $I_\ell$ of length $\ell$,
\begin{equation}
Pr\bigl\{ \forall [t_0,t] \subseteq I_\ell :\ 
\Aout(t_0,t) 
\geq \Ain\conv_{t_0} \T^{\ell,\eps}(t) \bigr\} \ge 1-\eps\  .  
\label{eq-eff-strongeffservice}
\end{equation}
This definition differs from the definition of an 
effective service curve
in Eqn.~(\ref{eq-eff-servicecurve}) in two ways: it uses the 
modified convolution operator, and it provides lower
bounds on the departures simultaneously in all 
subintervals of an interval $I_\ell$. 

With the strong effective adaptive network service curve, 
we obtain a probabilistic  version of a network service curve, 
with a similar concatenation formula as in the 
deterministic calculus. 
This is the content of the following theorem.

\begin{theorem}\ {\bf Concatenation of  Strong Effective 
Adaptive Service Curves.} 
\label{theorem-statnet-intervals}
Consider a flow that passes through $H$ network nodes in series. 
Assume that the functions $\T^{h,\ell,\eps}$ define 
strong effective adaptive service curves for intervals of length $\ell$ 
at each node  ($h=1,\dots , H$).  Then 
\begin{equation}
\T^{net, \ell, H \eps}(t) = \T^{1,\ell, \eps} \conv \ldots
                \conv \T^{H,\ell, \eps} (t)\ 
\label{eq-statnet-i1}
\end{equation}
is a strong effective adaptive service curve for intervals of length $\ell$.
\end{theorem}

Note the similarity of the convolution formula in 
Eqn.~(\ref{eq-statnet-i1}) with the corresponding expression 
in the deterministic calculus. 
Thus, in the statistical calculus, obtaining a 
statistical end-to-end service curve via a simple 
convolution operation comes at the price of significant modifications 
to the definition of a service curve.  

\medskip
\noindent{\bf Proof.} \ 
We need to show that $T^{net, \ell,H\eps}$ satisfies
for any interval $I_\ell$ of length $\ell$
\begin{equation}
Pr \bigl\{\forall [t_0,t]\subseteq I_\ell:\ 
 \Aout(t_0,t) \ge \Ain\conv_{t_0} \T^{net,\ell,H\eps}(t)
\bigr\} \ge 1-H\eps\ .
\end{equation}
The argument closely follows Steps 2 and 3 from
the proof of Theorem~\ref{theorem-statnet-intervals-x}.
%By assumption, we have the bounds
%\begin{equation}
%\label{eq-proof-strong-net-0} 
%Pr \bigl\{\forall [t_0,x]\subseteq I_\ell, \,\forall h=1,\dots H:\ 
% \Aout^h(t_0,x) \ge \Ain^h\conv_{t_0} \T^{h,\ell,\eps}(x)
%\bigr\} \ge 1-H\eps\ .
%\end{equation}
If, for a particular sample path, 
\begin{equation}
\label{eq-proof-strong-net-1} 
\forall [t_0,x]\subseteq I_\ell,\ \forall h=1,\dots , H:\quad 
\Aout^{h}(t_0,x) \ge \Ain^h \conv_{t_0}\T^{h,\ell,\eps}(x) \ ,
\end{equation}
then, for any fixed $[t_0,t]\subseteq I_\ell$,
the time-shifted arrivals
and departures defined by Eqn.~(\ref{eq-timeshift})
satisfy
\begin{equation}
\label{eq-proof-strong-net-2} 
\forall x\le t-t_0,\ \forall h=1,\dots , H:\quad 
\tilde \Aout^{h}(x) \ge \tilde \Ain^h \conv\T^{h,\ell,\eps}(x) \ .
\end{equation}
By Step~2 of proof of 
Theorem~\ref{theorem-statnet-intervals-x}, this implies 
\begin{equation}
\tilde \Aout(t) 
\ge \tilde \Ain \conv \bigl(\T^{1,\ell,\eps} \conv \dots \conv 
             \T^{H,\ell,\eps}\bigr) (t)\ .
\end{equation}
Reversing the time shift and using that $[t_0,t] \subseteq I_\ell$ was 
arbitrary, we arrive at
\begin{equation}
\label{eq-proof-strong-net-5} 
\forall [t_0,t] \subseteq I_\ell : \ 
\Aout(t) \ge \Ain \conv_{t_0} \bigl(\T^{1,\ell,\eps} 
                     \conv \dots \conv \T^{H,\ell,\eps}\bigr) (t)\ .
\end{equation}
We conclude that 
\begin{eqnarray}
&& Pr \bigl\{\forall [t_0,t]\subseteq I_\ell:\ 
 \Aout(t_0,t) \ge \Ain\conv_{t_0} \T^{net,\ell,H\eps}(t)
\bigr\} \\
&&\qquad \ge 
Pr \bigl\{\forall [t_0,x]\subseteq I_\ell, \,\forall h=1,\dots H:\ 
 \Aout^h(t_0,x) \ge \Ain^h\conv_{t_0} \T^{h,\ell,\eps}(x) \bigr\} \\ 
&&\qquad \ge 1-H\eps\ .
\end{eqnarray}
The first inequality follows from the
definition of  $\T^{net,\ell, H\eps}$ and the 
fact that Eqn.~(\ref{eq-proof-strong-net-1}) implies
Eqn.~(\ref{eq-proof-strong-net-5}), and the second inequality
uses the defining property  of strong effective $\ell$-adaptive
service curves.
\hfill$\Box$

\bigskip 
A comparison of the definition of the strong effective adaptive service curve 
in Eqn.~(\ref{eq-eff-strongeffservice}) with 
Eqn.~(\ref{eq-eff-intervalservice}) shows that a 
strong effective adaptive service curve is an effective $\ell$-adaptive 
service curve which provides service guarantees
simultaneously on all subintervals of an interval of length $\ell$.
A comparison of Theorem~\ref{theorem-statnet-intervals} 
with Theorems~\ref{theorem-statnet-intervals-x}, 
\ref{theorem-statnet-intervals-w}, 
and ~\ref{theorem-statnet-intervals-y}
shows that the more stringent strong effective adaptive service curve 
expresses the statistical calculus more concisely.
Therefore, unless additional assumptions are made 
on the arrival processes and the service curves, the network calculus 
with strong effective adaptive service curves offers the preferred framework. 

Our next result shows how to construct a strong 
effective adaptive service curve from an effective adaptive service curve. 
The lemma indicates that the choice of working with 
a strong effective adaptive service curve rather than an effective 
adaptive service curve is purely a matter of technical convenience.
\begin{lemma} 
\label{lemma-construct-strong}
If $\S^{\ell,\eps}$ is a nondecreasing function
which defines an effective $\ell$-adaptive
service curve for a flow,
then, for any choice of $a>0$, the function
\begin{equation}
\T^{\ell,\eps'}= \S^{\ell,\eps}\conv \delta_{a}\
\end{equation}
is a strong effective service curve for intervals
of length $\ell$, with violation probability given by 
\begin{equation}
\eps' = \lceil 2\ell/a \rceil^2 \eps/2\ .
\end{equation}
\end{lemma}

\medskip
\noindent{\bf Proof.} \ 
\bigskip
We will show that for any interval $I_\ell$ of length $\ell$,
\begin{equation}
\forall [t_0,t]\subseteq I_\ell:\quad 
Pr\bigl\{ \Aout(t_0,t)\ge \Ain\conv_{t_0}\S^{\ell,\eps}(t)\bigr\}\ge 1-\eps
\end{equation}
implies
\begin{equation}
Pr\bigl\{ \forall [t_0,t]\subseteq I_\ell:\ 
\Aout(t_0,t)\ge \Ain\conv_{t_0}\T^{\ell,{\eps'}}(t)\bigr\}\ge 1-\eps'\ ,
\end{equation}
where $\eps'$ and $\T^{\ell,{\eps'}}$ are as given in the statement of 
the lemma.  By performing a suitable time shift as  in 
Eqn.~(\ref{eq-timeshift}),
we may assume without loss of generality that $I_\ell=[0,\ell]$.  

The strategy is similar to the construction 
of strong effective envelopes from effective
envelopes in~\cite{Boorstyn2000b}, and uses the
same techniques as the first step in the proof
of Theorem~\ref{theorem-statnet-intervals-x}.
We first use the fact that the departures satisfy the
positivity assumption (A1) to translate service guarantees
given on a subinterval into a service
guarantee on a longer interval. 
In the second step, we establish probabilistic
bounds for the departures simultaneously
in a finite number of subintervals $I_{ij}$ of $I_\ell$,
and then bound the departures in
general subintervals of $I_\ell$ from below in terms of
the departures in  the $I_{ij}$.

\bigskip\noindent{\em Step 1: A property of the modified convolution.}  \ 
Let $g$ be a nondecreasing function, 
let $[t_1, t_2]\subseteq [t_0,t]$,
and $a\ge (t-t_0)-(t_2-t_1)$.
Then
\begin{equation}
\label{eq-proof-convert-1}
\Aout(t_1,t_2)\ge \Ain\conv_{t_1} g (t_2)
\end{equation}
implies
\begin{equation}
\label{eq-proof-convert-2}
\Aout(t_0,t)\ge \Ain\conv_{t_0} \bigl(g\conv\delta_a) (t)\ .
\end{equation}
To see this, note that Eqn.~(\ref{eq-proof-convert-1})
implies that either
\begin{equation}
\label{eq-proof-convert-3}
\Aout(t_1,t_2) \ge g(t_2-t_1)\ ,
\end{equation}
or
\begin{equation}
\label{eq-proof-convert-4}
\Aout(t_1,t_2) \ge B(t_1) +\inf_{\tau\in [t_1,t_2]} 
                 \{\Ain(t_1,\tau)+g(t_2-\tau)\}\ .
\end{equation}
If Eqn.~(\ref{eq-proof-convert-3}) holds, then
\begin{equation}
\Aout(t_0,t) \ge \Aout(t_1,t_2) \ge g(t_2-t_1) \ge g\conv 
\delta_a (t-t_0) \ge \Ain\conv_{t_0} \bigl(g\conv\delta_a) (t)\ ,
\end{equation}
proving Eqn.~(\ref{eq-proof-convert-2}) in this case. We have used that
$g$ is nondecreasing in the last inequality.
If Eqn~(\ref{eq-proof-convert-4}) holds, then
\begin{eqnarray}
\Aout(t_0,t) 
&\ge& \Aout(t_0,t_1) + B(t_1) +\inf_{\tau\in [t_1,t_2]} 
           \{\Ain(t_1,\tau)+g(t_2-\tau)\}
\label{eq-proof-convert-5}\\
&=& B(t_0) +\inf_{\tau\in [t_1,t_2]} 
           \{\Ain(t_0,\tau)+g(t_2-\tau)\}
\label{eq-proof-convert-6}\\
&\ge & B(t_0) +\inf_{\tau\in [t_0,t]} 
           \{\Ain(t_0,\tau)+\bigl(g\conv\delta_a)(t_2-\tau)\}
\label{eq-proof-convert-7}\\
&\ge & \Ain\conv_{t_0} \bigl(g\conv\delta_a) (t)\ .
\end{eqnarray}
which  proves Eqn.~(\ref{eq-proof-convert-2}) in the
second case. In Eqn.~(\ref{eq-proof-convert-5}) we have
used  Eqn.~(\ref{eq-proof-convert-4}). In Eqn.~(\ref{eq-proof-convert-6}),
we have used that $\Aout(t_0,t_1)+B(t_1) = B(t_0)+\Ain(t_0,t_1)$
and taken  $\Ain(t_0,t_1)$ under the infimum.
Eqn.~(\ref{eq-proof-convert-7})
uses the monotonicity of $g$ and extends the range of the infimum.

\medskip\noindent{\em Step 2: Uniform probabilistic bounds on $I_\ell$.} \ 
Fix $a>0$, set $x_i=i\, a/2$, and consider the intervals
\begin{equation}
I_{ij}= [x_i, x_j]\ , \quad 0\le i<j < n\ ,
\end{equation}
where  $n=\lceil 2\ell/a\rceil$ is 
the smallest integer no less than $2\ell/a$.
Consider the events
\begin{equation}
E_{ij} := \bigl\{ \Aout(x_i, x_j)\ge \Ain \conv_{x_i}\S^{\ell,\eps}(x_j)
       \bigr\}\ .
\end{equation}
%
%Since $\S^{\ell,\eps}$ is an effective $\ell$-adaptive
%service curve, we have that $Pr\{E_{ij}\}\ge 1-\eps$
%for each $i,j$,  and consequently the probability that
%all of the $E_{ij}$ occur simultaneously is bounded below by
%\begin{equation}
%\label{eq-construct-proof-main}
%Pr \Bigl\{ \bigcap_{0\le i<j< n} E_{ij} \Bigr\}
%\ge 1-n^2\eps/2\ .
%\end{equation}
Let $[t_0,t]\subseteq [0,\ell]$ be arbitrary, and
choose $I_{ij}\subseteq [t_0,t]$ be as large as possible.
If $E_{ij}$ occurs, we apply Step~1 with
$t_1=x_i$ and $t_2=x_j$, and use that $(x_i-t_0) + (t-x_j)\le a$
to see that
\begin{equation}
\Aout(t) \ge \Ain \conv_{t_0} \bigl(\S^{\ell,\eps}\conv\delta_a\bigr)(t)\ .
\end{equation}
It follows that
\begin{eqnarray}
\nonumber
&& Pr\bigl\{ \forall [t_0,t]\subseteq [0,\ell]:\ 
\Aout(t_0,t)\ge \Ain\conv_{t_0}
      \bigl(\S^{\ell,\eps}\conv\delta_a\bigr)(t)\bigr\}
\nonumber\\
&& \qquad \ge Pr \Bigl\{ \bigcap_{0\le i<j< n} E_{ij} \Bigr\}
\label{eq-construct-proof-2}\\
&&\qquad \ge 1- n^2\eps/2\ ,
\label{eq-construct-proof-3}
\end{eqnarray}
as claimed.  
Here,  Eqn.~(\ref{eq-construct-proof-2}) uses Step 2, and 
Eqn.~(\ref{eq-construct-proof-3}) uses the definition of $\S^{\ell,\eps}$.
\hfill$\Box$

\subsection{Recovering an effective service curve from 
effective adaptive service curves}

We next show that the adaptive versions of the effective 
service curve can yield effective service service curves 
in their original definition. This, however, requires us 
to add appropriate assumptions on the traffic at a node. 
The following lemma gives a sufficient conditions
for an effective $\ell$-adaptive curve to be an effective service curve
in the sense of Eqn.~(\ref{eq-eff-servicecurve}).
Combining Theorem~\ref{theorem-statnet-intervals-y} with 
Lemma~\ref{lemma-eff-sufficient-y} yields an effective 
network service curve, which by Theorem~\ref{theorem-statcalc} 
guarantees probabilistic bounds on output, backlog, and delay.

\begin{lemma}
\label{lemma-eff-sufficient-y}
Let $\S^{\ell,\eps}$ be a nondecreasing
function which defines an effective $\ell$-adaptive service curve 
for a flow with arrival process $A$, and an 
\begin{enumerate}
\item
If 
\begin{equation}
\label{eq-eff-sufficient-y1}
Pr \bigl\{ \exists t_0 \in [t-\ell,t]:\ B(t_0)=0 \bigr\}\ge 1-\eps_1
\end{equation}
for all $t >0$, then, for any choice of $a>0$, 
$\S^{\eps \ell/a +\eps_1}= \S^{\ell,\eps} \conv \delta_a$ 
is an effective adaptive service curve for intervals of arbitrary length,
with violation probability $\eps\ell/a+\eps_1$.
In particular,
\begin{equation}
\label{eq-eff-sufficient-y11}
Pr \bigl\{ D(t)\ge A*\S^{\ell,\eps}\conv\delta_a (t) \bigr\}
\ge 1- (\eps \ell /a +\eps_1)
\end{equation}
for all $t>0$.
\item
If the arrival process $A$ has arrival envelope $A^*$ and  
\begin{equation}
\label{eq-eff-sufficient-y2}
Pr \bigl\{ B(t)\le \S^{\ell,\eps} (\ell)-A^*(\ell) \bigr\} \ge 1- \eps_1\ ,
\end{equation}
for all $t \ge 0$, then $\S^{\eps+\eps_1} = \S^{\ell,\eps}$ is an effective  
adaptive service curve for intervals of arbitrary length, with violation
probability $\eps+\eps_1$.  
In particular, 
\begin{equation}
\label{eq-eff-sufficient-y22}
Pr \bigl\{ D(t)\ge A*\S^{\ell,\eps} (t) \bigr\}
\ge 1- (\eps+\eps_1) \ ,  
\end{equation}
for all $t \ge 0$. 
\end{enumerate}
\end{lemma}

The lemma should be
compared with Lemma~\ref{lemma-sufficient},
as both provide sufficient conditions under which
service guarantees on intervals of a given finite length
imply service guarantees on intervals of arbitrary length.
While   the condition on $\ell$ in Eqn.~(\ref{eq-def-T0}) involves only
the deterministic arrival envelope and the service curve,
Eqs.~(\ref{eq-eff-sufficient-y1}) and 
(\ref{eq-eff-sufficient-y2}) represent additional assumptions 
on the backlog process. 
This points out a fundamental difference between the 
deterministic and the statistical network calculus.

\medskip
\noindent{\bf Proof.} \ 
The proof consists of three steps. If Eqn.~(\ref{eq-eff-sufficient-y2})
holds,  the first step can be  omitted.
In the first step, we modify a given effective $\ell$-adaptive service 
curve to give uniform probabilistic lower bounds
on the departure of all intervals of the form
$[t_0,t]$, where $t$ is fixed and $t_0\in [t-\ell,t]$.
This is analogous to the
first step in the proof of Theorem~\ref{theorem-statnet-intervals-x}.
The second step contains a deterministic argument. We conclude with
a probabilistic estimate.

\medskip\noindent{\em Step 1: Uniform probabilistic bounds.} \ 
Suppose that $S^{\ell, \eps}$ is a nondecreasing effective
$\ell$-adaptive service  curve, that is,  for any $t\ge 0$,
\begin{equation}
\label{eq-stronger-5}
\forall t_0 \in [t-\ell,t]:\quad
Pr \bigl\{ \Aout(t_0,t)\ge \Ain\conv_{t_0}\S^{\ell,\eps}(t) \bigr\} 
\ge 1-\eps\ .
\end{equation}
We will show that then, for any choice of $a>0$,
\begin{equation}
\label{eq-stronger-6}
Pr \bigl\{ \forall t_0 \in [t-\ell,t]: 
\Aout(t_0,t)\ge \Ain\conv_{t_0}\S^{\ell,\eps}(t-a) \bigr\} \ge 1-\eps \ell/a\ .
\end{equation}
To see this, assume without loss of generality
that $t=\ell$, and consider the events
\begin{equation}
       E_j=\bigl\{ \Aout(x_j,t)\ge \Ain\conv_{x_j}\S^{\ell, \eps}(t) \bigr\}\ ,
\quad 0\le i<n\ .
\end{equation}
If $E_j$ occurs, we have for $x\in [x_{j-1},x_j)$ by the first
step of the proof of Lemma~\ref{lemma-construct-strong}
(with $t_0=x$, $t_1=x_j$, $t_2=t$), that
\begin{equation}
\Aout(x,t)\ge  \Ain\conv_{x_j}\bigl(\S^{\ell,\eps}\conv\delta_a\bigr)(t)\ .
\end{equation}
It follows that
\begin{eqnarray}
&& Pr \bigl\{ \forall t_0 \in [0,\ell]: 
\Aout(t)\ge \Ain\conv_{t_0}\S^{\ell, \eps}(t-a) \bigr\} 
\nonumber\\
%&&\qquad \ge Pr \bigl\{ \forall i=0,\dots , n-1: \ 
%\Aout(x_j)\ge \Ain\conv_{x_j}\S^{\ell, \eps}(x_j) \bigr\} \\
&&\qquad =  Pr\Bigl\{ \bigcap_{0 \le i < n} E_j\Bigr\}\\
&&\qquad \ge  1-n\eps\ ,
\end{eqnarray}
which proves Eqn.~(\ref{eq-stronger-6}) in the case $t=\ell$.

\medskip\noindent {\em Step 2: Deterministic argument.} \ 
Fix $t\ge 0$, and suppose that for a particular sample path, we have
\begin{equation}
\label{eq-proof-eff-suff-1a}
\forall x \in [t-\ell,t]:\ \Aout(x,t)\ge \Ain\conv_{x}
        \T^{\ell,\eps}(t)\ ,
\end{equation}
and either
\begin{eqnarray}
\label{eq-proof-eff-suff-2a}
1. && \exists t_0 \in [t-\ell,t]:
\ B(t_0)=0 \ , \mbox{or} \\
\label{eq-proof-eff-suff-2b}
2. && 
B(t-\ell)\le \S^{\ell,\eps} (\ell)-A^*(\ell) \ , 
\mbox{ where $A^*$ is an arrival envelope.}  
\end{eqnarray}
In the first case, we can set $x=t_0$ in
Eqn.~(\ref{eq-proof-eff-suff-1a}) to obtain
\begin{equation}
\label{eq-proof-eff-suff-3}
\Aout(t) \ge \inf_{\tau\le t-t_0} \{\Ain(t-\tau) + \S^{\ell,\eps}(\tau) \} .
\end{equation}
In the second case, we note 
that from Eqn.~(\ref{eq-proof-eff-suff-2b}) it follows that
\begin{equation}
B(t-\ell) + \inf_{\tau\le \ell} \{\Ain(t-\ell,t-\tau) + 
             \S^{\ell,\eps}(\tau) \}
\le \S^{\ell,\eps} (\ell)-A^*(\ell) + \Ain(t-\ell, t) 
\le \S^{\ell,\eps}(\ell) \ , 
\end{equation}
which implies that
\begin{equation}
\Ain\conv_{t-\ell}(t) = B(t - \ell)+ \inf_{\tau\le \ell}\{ \Ain(t-\ell,\tau)
        +\S^{\ell\eps}(\tau)\}\ .
\end{equation}
Inserting this into Eqn.~(\ref{eq-proof-eff-suff-1a})
with $x = t - \ell$ yields again Eqn.~(\ref{eq-proof-eff-suff-3}). 

\medskip\noindent{\em Step 3: Probabilistic estimate.\ }
If Eqn.~(\ref{eq-eff-sufficient-y1}) holds, we use Step 2 to see that
\begin{eqnarray}
&& 
Pr \bigl\{ \Aout(t)\ge \inf_{\tau\le \ell} \{ \Ain(t-\tau)+\Aout(\tau)\}
 \bigr\} \nonumber\\
&& \qquad \ge 
Pr\left\{
\begin{array}{l}
\forall x \in [t-\ell,t]:\ 
\Aout(x,t)\ge \Ain\conv_{x}\T^{\ell,\eps}(t), \ \mbox{and}\ 
\exists t_0 \in [t-\ell,t]: \ B(t_0)=0 
\end{array} \right\} \\
&& \qquad \ge 1-(\eps \ell/a+\eps_1)\ ,
\label{eq-eff-servicecurve-ell-1}
\end{eqnarray}
where we have used the result of Step 2 in the second line.
If in Eqn.~(\ref{eq-eff-sufficient-y2}) holds, we
have by Step 2
\begin{eqnarray}
&& \nonumber
Pr \bigl\{ \Aout(t)\ge \inf_{\tau\le \ell} \{ \Ain(t-\tau)+\Aout(\tau)\}
 \bigr\} \\
&& \qquad \ge 
Pr\left\{
\begin{array}{l}
\Aout(t-\ell,t)\ge \Ain\conv_{t-\ell}\T^{\ell,\eps}(t), \ \mbox{and}\ 
 B(t-\ell)\le \S^{\ell,\eps}(\ell)-A^*(\ell)
\end{array}
\right\} \\
&& \qquad \ge 1-(\eps +\eps_1)\ ,
\label{eq-eff-servicecurve-ell-2}
\end{eqnarray}
where we have used the definition of $\S^{\ell,\eps}$
in the second line. 
\hfill$\Box$

\bigskip
Note that Eqs.~(\ref{eq-eff-servicecurve-ell-1}) 
and~(\ref{eq-eff-servicecurve-ell-2})
supply time-independent bounds
on the range of the convolution, of the form given 
in Eqn.~(\ref{eq-eff-servicecurve-T}).
A similar, but simpler result holds for strong effective
$\ell$-adaptive  service curves:

\begin{lemma} 
\label{lemma-eff-sufficient}
Given a flow with arrival process $A$, and a 
strong effective adaptive service curve $\T^{\ell,\eps}$ on intervals 
of length $\ell$. Assume that for every $t\ge 0$, 
either Eqn.~(\ref{eq-eff-sufficient-y1}) or 
or Eqn.~(\ref{eq-eff-sufficient-y2}) 
is satisfied.
Then, for any $t\geq 0$ and any $t_1\le t$,
\begin{equation}
Pr \bigl\{ D(t_1,t)\ge A\conv_{t_1}\T^{\ell,\eps} (t) \bigr\}
\ge 1- (\eps + \eps_1) \ .  
\end{equation}
In particular, for $t_1 = 0$, $\S^{\eps+\eps_1} = \T^{\ell,\eps}$ is 
an effective  service curve in the sense of 
Eqn.~(\ref{eq-eff-servicecurve}).  
\end{lemma}

\medskip
\noindent{\bf Proof.} \ 
We need to show that under the assumptions of the lemma,
we have for any $t>0$ and any $t_1\le t$
\begin{equation}
\label{eq-proof-eff-suff-1}
Pr \bigl\{ \Aout(t_1,t)\ge 
\Ain\conv_{t_1}\T^{\ell,\eps}(t)\bigr\}\ge 1-(\eps+\eps_1)\ .
\end{equation}
By considering time-shifted arrivals
and departures as in Eqn.~(\ref{eq-timeshift}),
we may assume without loss of generality that $t_1=0$.
The first step in the proof of Lemma~\ref{lemma-eff-sufficient-y}
shows that
\begin{eqnarray}
&& 
Pr \bigl\{ \Aout(t)\ge \inf_{\tau\le \ell} \{ \Ain(t-\tau)+\Aout(\tau)\}
 \bigr\} 
\nonumber\\
&& \qquad \ge 
Pr\left\{
\begin{array}{l}
\forall [x,y]\subset [t-\ell,t]:\ 
\Aout(x,y)\ge \Ain\conv_{x}\T^{\ell,\eps}(x), \ \\
\mbox{and}\ \bigl\{ \mbox{ either } \ \exists t_0 \in [t-\ell,t]: \ B(t_0)=0 \ \\ 
\qquad \mbox{or} \ B(t-\ell)\le \T^{\ell,\eps} (\ell)-A^*(\ell)  \bigr\}
\end{array}
\right\} 
\label{eq-proof-eff-sufficient-1}\\
&& \qquad \ge 1-(\eps+\eps_1)\ ,
\label{eq-proof-eff-sufficient-2}
\end{eqnarray}
as claimed.  
\hfill$\Box$

\section{Conclusions}
\label{sec-concl}

We have presented a network calculus with probabilistic service guarantees 
where arrivals to the network satisfy a deterministic arrival bound. 
We have introduced the notion of {\em effective service curves} as a 
probabilistic bound on the service received by individual flows in 
a network. 
We have shown that some key results from the deterministic network calculus 
can be carried over to the statistical framework
by inserting appropriate probabilistic arguments. 

We showed that the deterministic  bounds on output, delay, and backlog from 
Theorem~\ref{theorem-detcalc} have corresponding formulations 
in the statistical calculus (Theorem~\ref{theorem-statcalc}).
We have extended the concatenation formula of
Theorem~\ref{theorem-detnet} for network
service curves to a statistical setting 
(Theorems~\ref{theorem-statnet-intervals-x}, 
~\ref{theorem-statnet-intervals-w}, 
\ref{theorem-statnet-intervals-y}, and ~\ref{theorem-statnet-intervals}). 
We showed that a modified effective service 
curve, called {\em strong effective adaptive service curve} yields 
the simplest concatenation formula.
In order to connect the different notions of effective service 
curves, we have made an additional assumption on the backlog in
Lemmas~\ref{lemma-eff-sufficient-y} and ~\ref{lemma-eff-sufficient}.
The results in this paper showed that a multi-node version of the 
statistical network calculus requires us to make assumptions 
that limit the range of the convolution operation 
when concatenating effective service curves. 
Such limits on a `maximum relevant time scale', 
can follow from assumptions on the traffic load 
(as in Theorems~\ref{theorem-statnet-intervals-w}, 
Lemma~\ref{lemma-eff-sufficient-y} and Lemma ~\ref{lemma-eff-sufficient}), 
or from appropriately modified service curves. 
While the question is open whether one can dispense with 
these additional assumptions, we have made an attempt to justify 
the need for them.

\section*{Acknowledgments}
We thank Rene Cruz for pointing out problems in earlier versions 
of the effective network service curve. 
The authors gratefully acknowledge the valuable comments from Chengzhi Li.

\newpage

\appendix 
\begin{center}
{\LARGE \bf APPENDIX} 
\end{center}

\section{Proof of Lemma~\ref{lemma-sufficient}} 

Let $\Smin^\ell$ be an $\ell$-adaptive service curve.  We need to show that 
\begin{equation}
\Aout(t_1,t)\ge \Ain\conv_{t_1}\Smin^\ell(t)
\end{equation}
holds for all $t, t_1\ge 0$ with $t_1\le t$.
By considering the time-shifted arrivals and departures
as in Eqn.~(\ref{eq-timeshift}),
we may assume without loss of generality 
that $t_1=0$. 

Consider intervals $I_\ell^k=[k\ell,(k+1)\ell]$, where
$k\ge 0$ is an integer.  We will show by induction, that  for any
integer $k\ge 0$,
\begin{equation}
\forall t\in  I_\ell^k:\quad
\Aout(t)\ge \Ain\conv \Smin^\ell(t)\ .
\label{eq-suff-0}
\end{equation}
Applying the definition of an $\ell$-adaptive network service
curve with $I_\ell=[0,\ell]$, we see that 
Eqn.~(\ref{eq-suff-0}) clearly holds for $k=0$.

For the inductive step, suppose
that Eqn.~(\ref{eq-suff-0}) holds for
some integer $k\ge 0$.
Fix $t\in I_\ell^{k+1}$,
and let $t_0=t-\ell\in I_\ell^{k}$.
By the inductive assumption, 
$\Aout(t_0)\ge \Ain * \Smin^\ell(t_0)$.
Eqn.~(\ref{eq-def-intervalservice}) says that either
\begin{equation}
\label{eq-suff-1}
\Aout(t_0,t)\ge \Smin^\ell(t-t_0)
\end{equation}
or
\begin{equation}
\label{eq-suff-2}
\Aout(t_0,t)\ge B(t_0) + \inf_{\tau\le t-t_0} 
\bigl\{\Ain(t-t_0,t-\tau) + \Smin^\ell(\tau)\bigr\}\ .
\end{equation}
If Eqn.~(\ref{eq-suff-1}) holds, then
\begin{eqnarray}
\Aout(t) &=& \Aout(t_0,t) +\Aout(t_0)
\label{eq-suff-3}\\
   &\ge& \Smin^\ell(t-t_0) + \inf_{\tau\le t_0} 
        \bigl\{\Ain(t_0-\tau) + \Smin^\ell(\tau)\bigr\}
\label{eq-suff-4}\\ 
   &\ge& \Smin^\ell (t-t_0) - A^*(t-t_0) + 
             \inf_{\tau\le t_0} \bigl\{\Ain(t-\tau) + \Smin^\ell(\tau) \bigr\}
\label{eq-suff-5}\\ 
   &\ge& \Ain \conv \Smin^\ell(t)\ .
\label{eq-suff-6}
\end{eqnarray}
In Eqn.~(\ref{eq-suff-4}), we have used Eqn.~(\ref{eq-suff-1})
and the inductive assumption.  In Eqn.~(\ref{eq-suff-5}), 
we have used that $\Ain(t_0-\tau,t-\tau)\le A^*(t-t_0)$ and pulled $A^*(t-t_0)$
out  of the infimum.
In Eqn.~(\ref{eq-suff-6}), we have inserted $t-t_0=\ell$,
used the assumption that $A^*(\ell) \le \Smin^{\ell}(\ell)$,
and extended the range of the infimum. 

If Eqn.~(\ref{eq-suff-2}) holds, then
\begin{eqnarray}
\Aout(t)  &=& \Aout(t_0)+\Aout(t-t_0) \\
&\ge&  \Ain (t_0) + \inf_{\tau\le t-t_0} 
            \bigl\{\Ain(t_0,t-\tau)+\Smin^\ell(\tau)\bigr\}
\label{eq-suff-7}\\
    &=&  \inf_{\tau\le t-t_0} \bigl\{ \Ain(t-\tau)+\Smin^\ell(\tau) \bigr\}
\label{eq-suff-8}\\
    &\ge& \Ain \conv \Smin^\ell(t)\ . 
\label{eq-suff-9}
\end{eqnarray} 
In Eqn.~(\ref{eq-suff-7}), we have 
used Eqn.~(\ref{eq-suff-2}), and
the fact that $\Aout(t_0)+B(t_0)=\Ain(t_0)$.
In Eqn.~(\ref{eq-suff-8}), 
we have taken $\Ain(t_0)$ under the infimum
and  used that $\Ain(t_0)+\Ain(t_0,t-\tau)=\Ain(t-\tau)$.
In Eqn.~(\ref{eq-suff-9}), we have extended the range of the infimum
and used the definition of the convolution.

Since $t\in I_\ell^{k+1}$ was arbitrary, this proves the 
inductive step, and the lemma.


\begin{thebibliography}{10}

\bibitem{Cruz99}
R.~Agrawal, R.~L. Cruz, C.~Okino, and R.~Rajan.
\newblock Performance bounds for flow control protocols.
\newblock {\em {IEEE/ACM} Transactions on Networking}, 7(3):310--323, June
  1999.

\bibitem{andrews00}
M.~Andrews.
\newblock Probabilistic end-to-end delay bounds for earliest deadline first
  scheduling.
\newblock In {\em Proceedings of IEEE Infocom 2000}, pages 603--612, Tel Aviv,
  March 2000.

\bibitem{Book-Baccelli}
F.~L. Baccelli, G.~Cohen, G.~J. Olsder, and J.-P. Quadrat.
\newblock {\em Synchronization and Linearity: An Algebra for Discrete Event
  Systems}.
\newblock John Wiley and Sons, 1992.

\bibitem{Boorstyn2000b}
R.~Boorstyn, A.~Burchard, J.~Liebeherr, and C.~Oottamakorn.
\newblock Statistical service assurances for traffic scheduling algorithms.
\newblock {\em IEEE Journal on Selected Areas in Communications. Special Issue
  on Internet QoS}, 18(12):2651--2664, December 2000.

\bibitem{LeBoudec98}
J.~Y.~Le Boudec.
\newblock Application of network calculus to guaranteed service networks.
\newblock {\em {IEEE/ACM} Transactions on Information Theory},
  44(3):1087--1097, May 1998.

\bibitem{Book-LeBoudec}
J.~Y.~Le Boudec and P.~Thiran.
\newblock {\em Network Calculus}.
\newblock Springer Verlag, Lecture Notes in Computer Science, LNCS 2050, 2001.

\bibitem{chang94}
C.~S. Chang.
\newblock Stability, queue length, and delay of deterministic and stochastic
  queueing networks.
\newblock {\em {IEEE} Transactions on Automatic Control}, 39(5):913--931, May
  1994.

\bibitem{Chang98a}
C.~S. Chang.
\newblock On deterministic traffic regulation and service guarantees: a
  systematic approach by filtering.
\newblock {\em {IEEE/ACM} Transactions on Information Theory},
  44(3):1097--1110, May 1998.

\bibitem{Book-Chang}
C.~S. Chang.
\newblock {\em Performance Guarantees in Communication Networks}.
\newblock Springer Verlag, 2000.

\bibitem{Cruz91a}
R.~L. Cruz.
\newblock A calculus for network delay, {Part I} : Network elements in
  isolation.
\newblock {\em {IEEE} Transaction of Information Theory}, 37(1):114--121, 1991.

\bibitem{Cruz91b}
R.~L. Cruz.
\newblock A calculus for network delay, {Part II} : Network analysis.
\newblock {\em IEEE Transactions on Information Theory}, 37(1):132--141,
  January 1991.

\bibitem{Cruz95}
R.~L. Cruz.
\newblock Quality of service guarantees in virtual circuit switched networks.
\newblock {\em {IEEE} Journal on Selected Areas in Communications},
  13(6):1048--1056, August 1995.

\bibitem{Cruz96a}
R.~L. Cruz.
\newblock Quality of service management in integrated services networks.
\newblock In {\em Proceedings of the 1st Semi-Annual Research Review, CWC},
  UCSD, June 1996.

\bibitem{Cruz96b}
R.~L. Cruz and C.~Okino.
\newblock Service gurantees for flow control protocols.
\newblock In {\em Proceedings of the 34th Allerton Conference on
  Communications, Control and Computating}, October 1996.

\bibitem{ElMi99}
A.~Elwalid and D.~Mitra.
\newblock Design of generalized processor sharing schedulers which
  statistically multiplex heterogeneous {QoS} classes.
\newblock In {\em Proceedings of {IEEE} {INFOCOM}'99}, pages 1220--1230, New
  York, March 1999.

\bibitem{Kurose92}
J.~Kurose.
\newblock On computing per-session performance bounds in high-speed multi-hop
  computer networks.
\newblock In {\em {ACM} {Sigmetrics}'92}, pages 128--139, 1992.

\bibitem{knightly2000b}
C.~Li and E.~Knightly.
\newblock Coordinated network scheduling: A framework for end-to-end services.
\newblock In {\em Proceedings of IEEE ICNP 2000}, Osaka, November 2000.

\bibitem{Knightly99b}
J.~Qiu and E.~Knightly.
\newblock Inter-class resource sharing using statistical service envelopes.
\newblock In {\em Proceedings of IEEE Infocom '99}, pages 36--42, March 1999.

\bibitem{Chiussi99}
V.~Sivaraman and F.~M. Chiussi.
\newblock Statistical analysis of delay bound violations at an earliest
  deadline first scheduler.
\newblock {\em Performance Evaluation}, 36(1):457--470, 1999.

\bibitem{Chiussi2000}
V.~Sivaraman and F.~M. Chiussi.
\newblock Providing end-to-end statistical delay guarantees with earliest
  deadline first scheduling and per-hop traffic shaping.
\newblock In {\em Proceedings of IEEE Infocom 2000}, pages 603--612, Tel Aviv,
  March 2000.

\bibitem{StaSi00}
D.~Starobinski and M.~Sidi.
\newblock Stochastically bounded burstiness for communication networks.
\newblock {\em {IEEE} Transaction of Information Theory}, 46(1):206--212, 2000.

\bibitem{Yaron93}
O.~Yaron and M.~Sidi.
\newblock Performance and stability of communication networks via robust
  exponential bounds.
\newblock {\em {IEEE/ACM} Transactions on Networking}, 1(3):372--385, June
  1993.

\end{thebibliography}
\end{document}